\documentclass[12pt,a4paper,oneside]{article}

\usepackage{mathpazo}
\usepackage{mathtools}
\usepackage{textcomp}
\usepackage{siunitx}
\usepackage{hyperref}
\usepackage{graphicx}
\usepackage{xcolor}
\usepackage{xargs} 
\usepackage[colorinlistoftodos,prependcaption,textsize=tiny]{todonotes}
\usepackage{setspace}
\usepackage{placeins}
\usepackage[italic]{hepnames}
\usepackage[export]{adjustbox}
\usepackage{morefloats}
\usepackage{afterpage}
\usepackage{titling}
\usepackage{geometry}
\usepackage{lineno}

\title{LYCORIS - A Large Area Strip Telescope}
\author{
  Uwe Kr\"amer \thanks{DESY Hamburg} , Dimitra Tsionou *, Marcel Stanitzki *, Mengqing Wu *
  }

\begin{document}

	\maketitle
	\begin{center}
		Talk presented at the International Workshop on Future Linear Colliders (LCWS2017), Strasbourg, France, 23-27 October 2017. C17-10-23.2.
	\end{center}
	\begin{abstract}
		The LYCORIS Large Area Silicon Strip Telescope for the DESY II Test Beam Facility is presented. The  DESY II Test Beam Facility provides electron and positron beams for beam tests of up to $\SI{6}{GeV}$. A new telescope with a large $10\times \SI{20}{cm^2}$ coverage area based on a $\SI{25}{\micro m}$ pitch strip sensor is to be installed within the PCMAG $\SI{1}{T}$ solenoid. The current state of the system is presented. 
	\end{abstract}

	\section{Introduction}\label{sec:Introduction}
		\begin{figure} 
			\begin{center}
				\includegraphics[width=0.9 \textwidth]{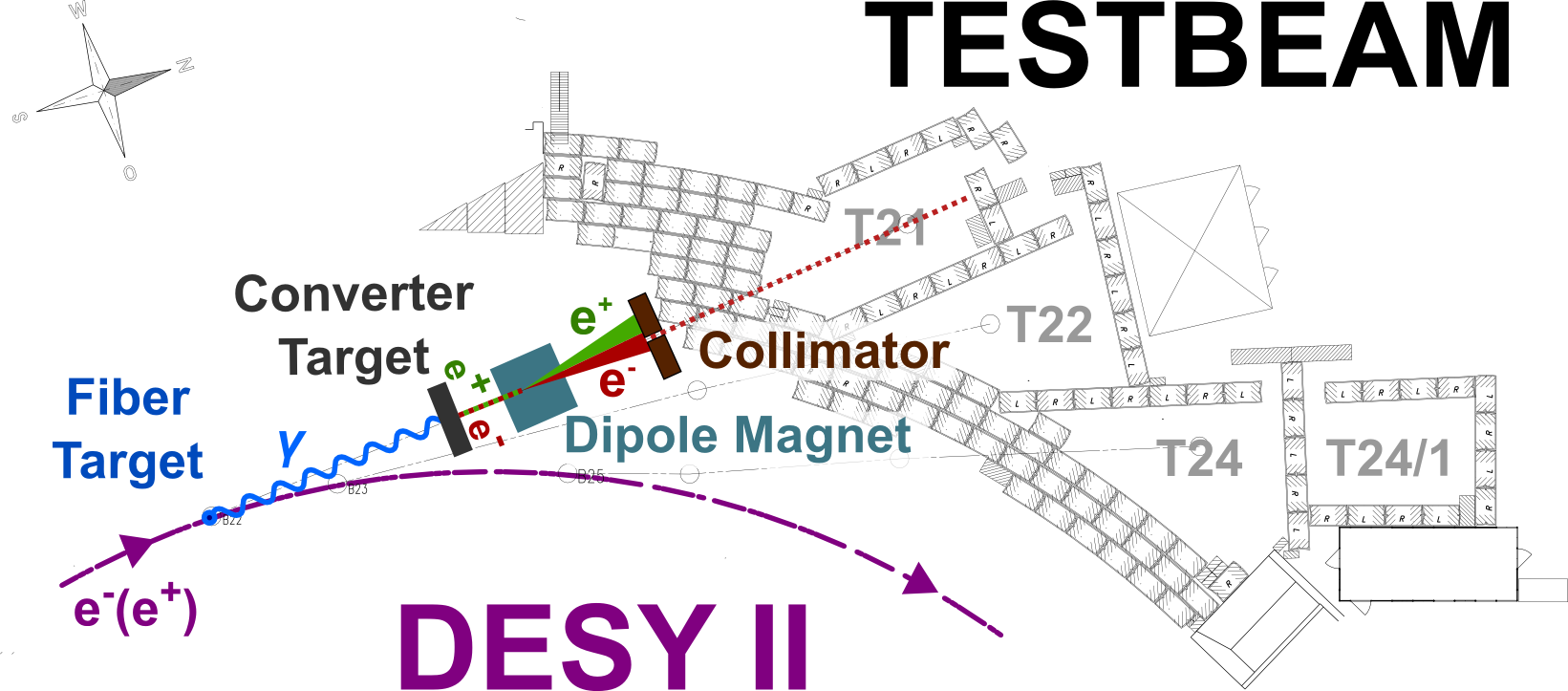}
				\caption{Schematic layout of the DESY II Test Beam Facility.\cite{testbeamwebsite}} 
				\label{fig:desy_tb}
			\end{center}
		\end{figure}	
		The DESY II Test Beam Facility, see see Figure \ref{fig:desy_tb}, provides an electron/positron beam with energies of up to $\SI{6}{GeV}$ to perform beam tests. Beam tests are an essential tool for detector development in high energy particle physics, photon science and heavy ion physics.
		Primary electrons from the DESY II synchrotron hit a fiber target placed within the beam pipe to produce Bremsstrahlung photons that are then converted into $\mathrm{e}^+ \mathrm{e}^-$ pairs by hitting a converter target. The particle type (electrons/positrons) and energies (0.6 to $\SI{6.0}{GeV}$) can be chosen using an adjustable dipole magnet and adjustable collimator. From the collimator the selected particles travel through a shutter to the respective beam area.
		
\begin{figure} 
			\begin{center}
				\includegraphics[scale=0.12]{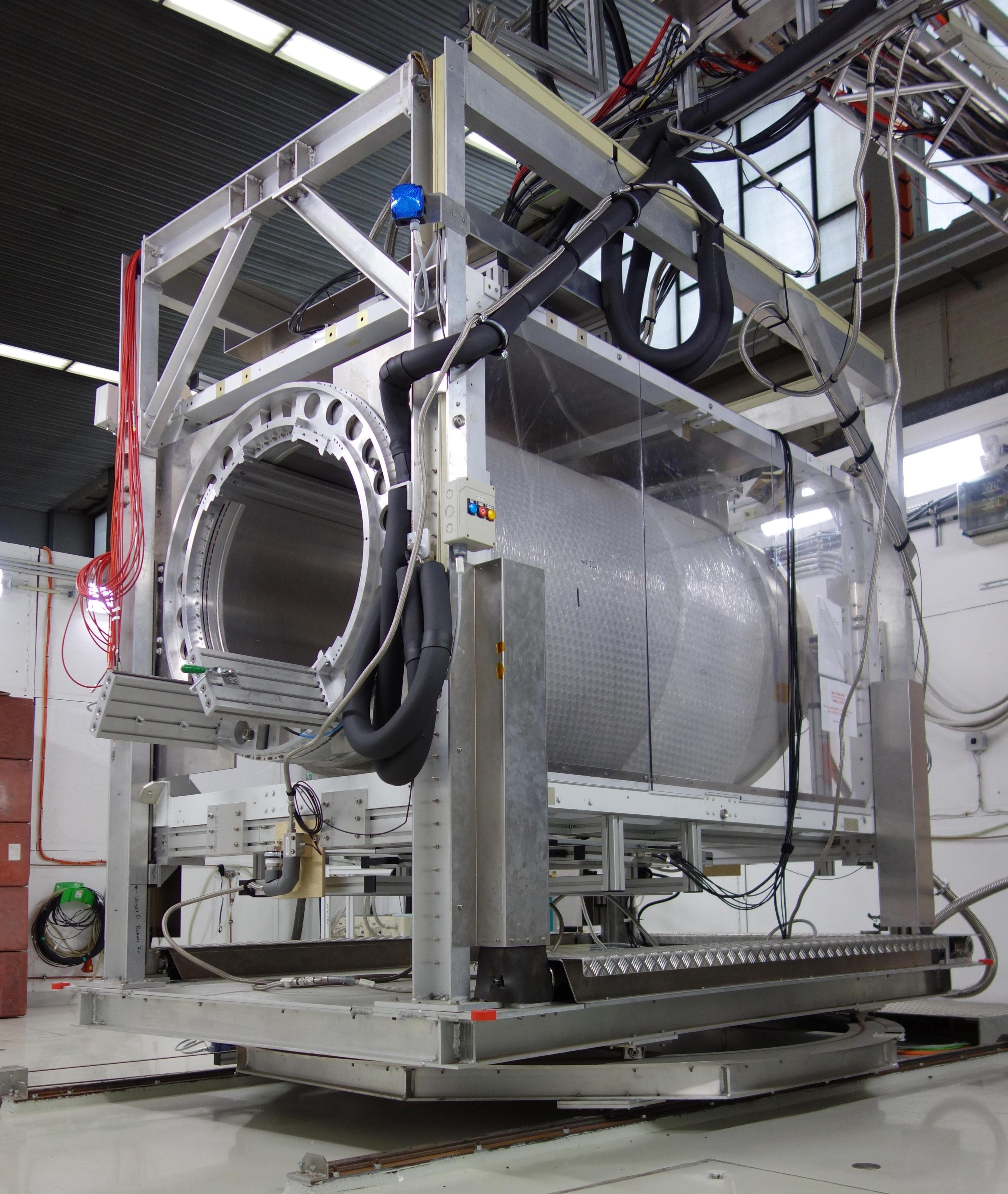}
				\caption{The PCMAG 1T solenoid mounted on the movable stage at DESY II test team area T24/1.}
				\label{fig:pcmag}
			\end{center}
		\end{figure}

		A selection of infrastructures have been installed in the beam lines. This includes the superconducting $\SI{1}{T}$ solenoid (PCMAG). The PCMAG is mounted on a stage that can be moved along- and rotated around three axes. It possesses an inner usable diameter of $\SI{75}{cm}$ and is capable of providing a $\SI{1}{T}$ magnetic field perpendicular to the beam axis, see Figure \ref{fig:pcmag}\cite{pcmag:magnet}\cite{pcmag:fieldmeas}\cite{pcmag:fieldana}.  
		Other infrastructure at the DESY II Test Beam Facility includes the DATURA/DURANTA silicon telescopes. Each telescope consists of 6 layers of the Mimosa26 chip with an active area of $1 \times \SI{2}{cm^2}$ and a pitch of $\SI{18}{\micro m}$ resulting in a $\SI{3}{\micro m}$ tracking resolution \cite{EudetTel2016}. Each sensor is encased in an aluminum cassette and cooled via water cooling. The synchronization of the telescopes with the Device Under Test (DUT) is done by using the EUDET Trigger Logic Unit (TLU) \cite{EudetTel2016} \cite{EUDETMemo2009}. Although these telescopes provide an excellent tracking resolution, the large support structure and small active area make them unsuitable for scanning a large area or when there are very tight constraints on the available space. 
		Therefore, a new silicon strip telescope possessing a large coverage area with a small support structure is being designed.
		The system requirements presented here are for the use case of a Time Projection Chamber (TPC) \cite{tpc2004} within the PCMAG as one of the most demanding scenario. 
		
\begin{figure} 
			\begin{center}
				\includegraphics[scale=0.22]{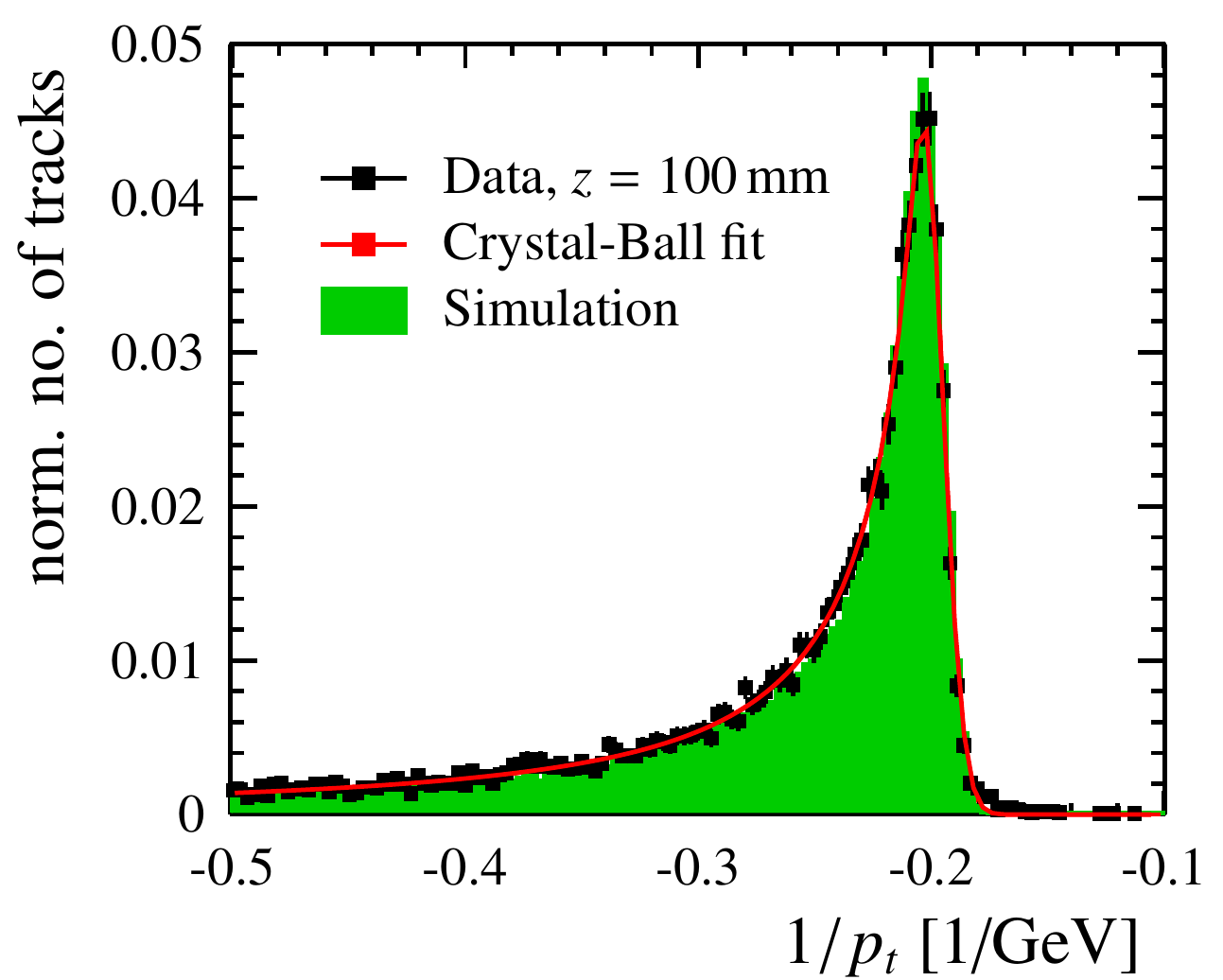}
				\caption{Momentum distribution of test beam particles after interactions with the PCMAG wall.  \cite{Mueller:2016exq}}
				\label{fig:moment_spread}
			\end{center}
		\end{figure}

		The telescope is used to provide a precise reference of the particle trajectory that can be used to study and correct for potential inhomogeneities of the electric field within the TPC volume, limiting the achievable momentum resolution. In addition, as a result of interactions of the test beam particles with the magnet wall, the test beam particles have a large spread in momentum when entering the magnet volume as seen in Figure \ref{fig:moment_spread}. The installation of a reference tracker to measure the curvature of tracks within the magnet, would provide a precise measurement of the test beam particle momentum after their interaction with the magnet wall. This could then be used to determine the achievable momentum resolution of the TPC under test.
		
		To meet these criteria, the telescope needs to provide a spatial single point resolution of better than $\sigma_{\mathrm{bend}} = \SI{10}{\micro m}$ along the bending direction within the PCMAG and a resolution better than $\sigma_{\mathrm{drift}} = \SI{1}{mm}$ along the drift axis of the TPC. These requirements stem from studies of the LCTPC collaboration. Other user groups interested in a large area reference tracker include, for example, the CALICE collaboration and the BELLE II collaboration.
		\FloatBarrier
	\section{The Telescope Sensors} \label{sec:sensors}
\begin{figure} 
			\begin{center}
				\includegraphics[scale=0.25]{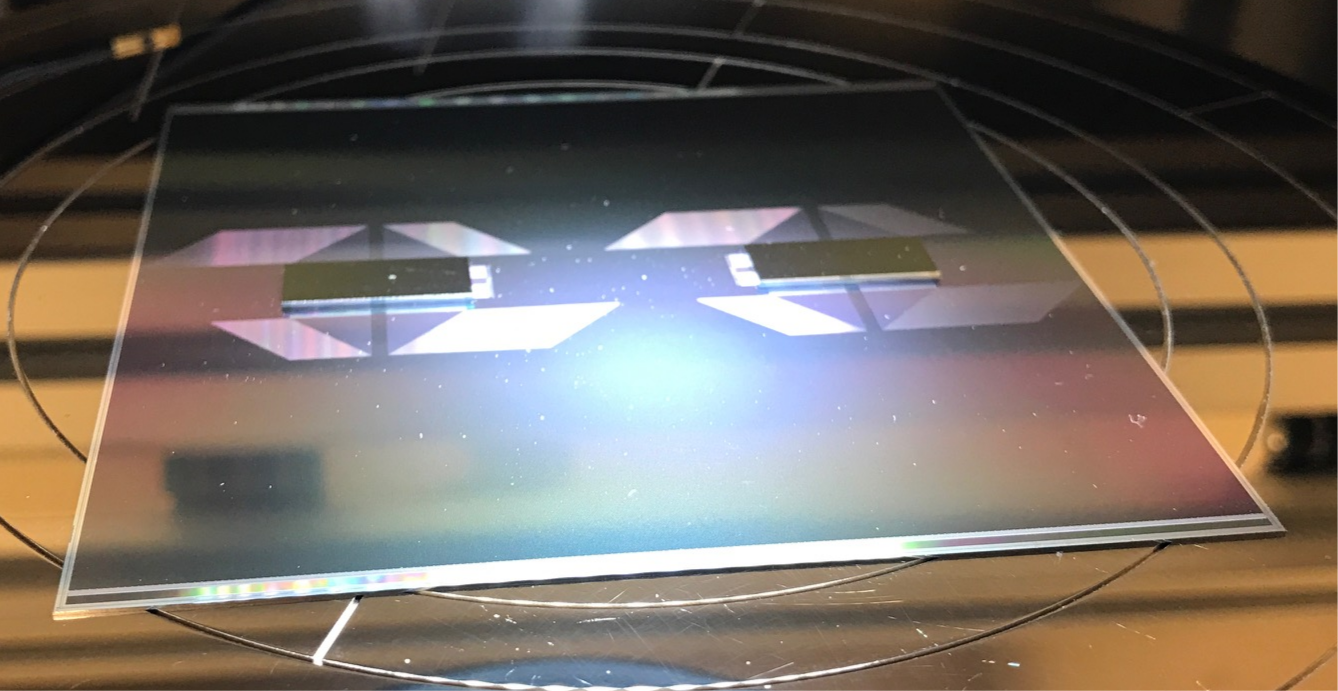}
				\caption{The SiD silicon strip sensor with two bump bonded KPiX chips on the clean room probe station.}
				\label{fig:strip_sensor}
			\end{center}
		\end{figure}

		Based on the requirements listed in section \ref{sec:Introduction}, the Silicon Detector (SiD) \cite{ILD2013} large area silicon strip sensors (Figure: \ref{fig:strip_sensor}) were selected for this beam telescope. These sensors were originally designed by SLAC for the International Linear Collider (ILC) \cite{ILC2013}. They possess an active area of $10 \times \SI{10}{cm^2}$ with a strip pitch of $\SI{25}{\micro m}$ and a radiation length of $X_0 = \SI{0.3}{\percent}$. The sensors will be read out with an integrated digital readout chip called KPiX that is bump bonded directly onto the sensor and routed to the strips within the silicon. A more detailed description of the KPiX readout chip can be found in section \ref{sec:kpix}). 
		
		29 sensors were delivered by Hamamatsu in August 2017 and subsequently tested at DESY for their electrical properties before they were sent off to Fraunhofer IZM for bump bonding of the KPiX readout chips onto the sensors. 
		The electrical properties of the sensors were tested by looking at the current over voltage (IV) and capacitance over voltage (CV) response of the sensor before and after the bump bonding. This was done at DESY using the probe station shown in Figure \ref{fig:strip_sensor}. High voltage from 0 to $\SI{300}{V}$ was applied to the backside of the sensor to deplete the sensor with probes located at the bias ring of the sensor. The current and capacitance of the sensor were measured.
		
		\begin{figure} 
			\begin{center}
			\includegraphics[scale=0.32]{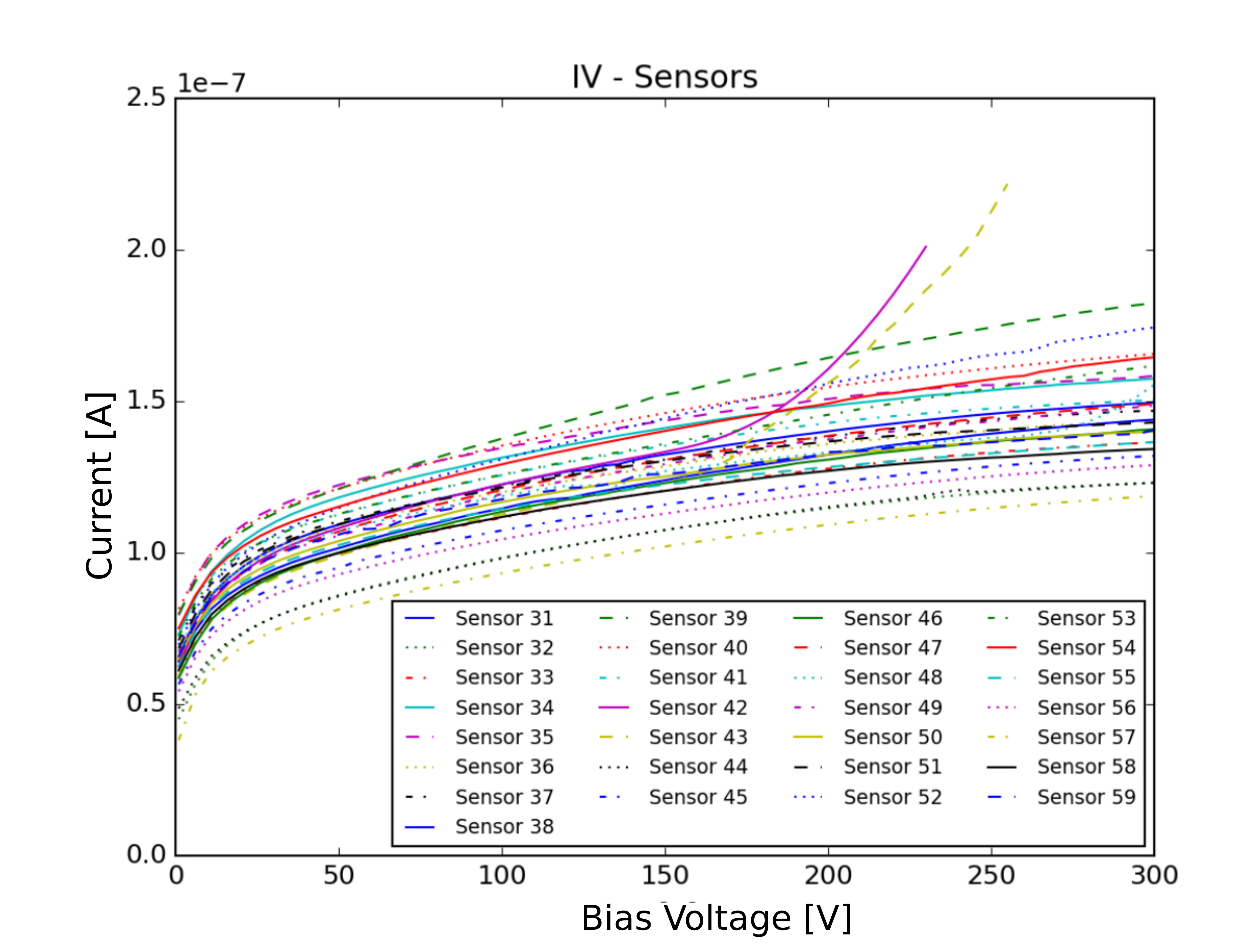}
			\caption{Current over bias Voltage of all 29 sensors. Sensors 42 and 43 show the beginning of a breakdown at higher voltages.}
				\label{fig:sensor_IV}
			\end{center}
		\end{figure}
		\begin{figure} 
			\begin{center}
				\includegraphics[scale=0.32]{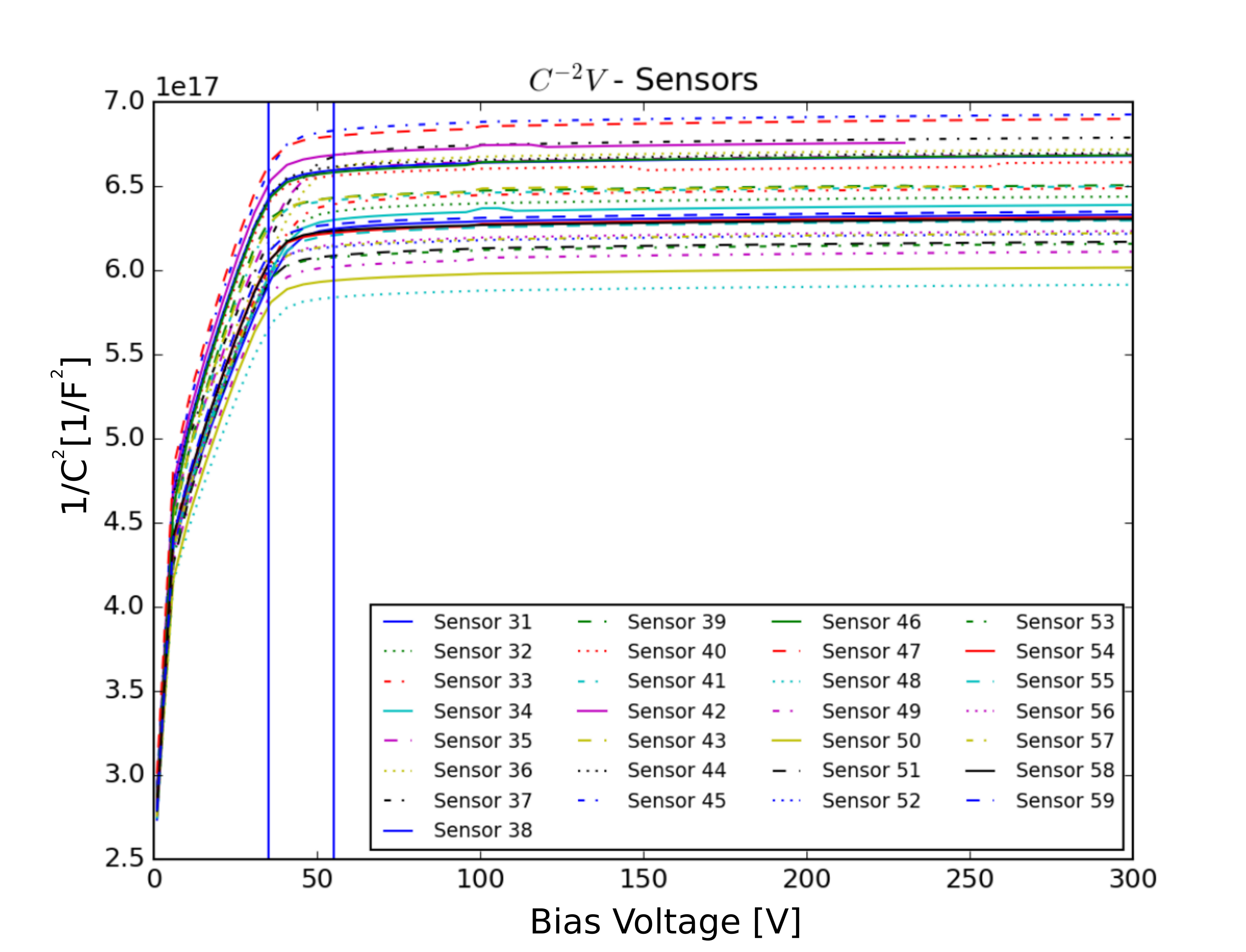}
				\caption{$C^{-2}$ over bias Voltage of all 29 sensors. The flat area shows the full depletion of the sensor beginning at around $\SI{55}{V}$}
				\label{fig:sensor_CV}
			\end{center}
		\end{figure}
		The IV measurements in Figure \ref{fig:sensor_IV} show all sensors to possess a dark current of around $I_{\mathrm{dark}} = \SI{100}{nA}$. Sensors 42 and 43 show the beginning of a breakdown at a bias voltage of around $V \approx \SI{280}{V}$. The CV measurements, see Figure \ref{fig:sensor_CV}, show that the sensors fully deplete at $V_{\mathrm{depletion}} \approx \SI{55}{V}$. As a result, the breakdown shown for sensors 42 and 43 are only of minor importance for use at the DESY II Test Beam Facility, as the sensors will not be subjected to significant doses of radiation which has been shown to increase the voltage needed for full depletion of the sensors \cite{VERBITSKAYA200347}.
		As all sensors show the expected low current and low depletion voltage, all 29 sensors were sent to IZM for bump bonding to the KPiX readout chips.
		\FloatBarrier
	\section{KPiX Readout Chip} \label{sec:kpix}
\begin{figure} 
			\begin{center}
				\includegraphics[scale=0.5]{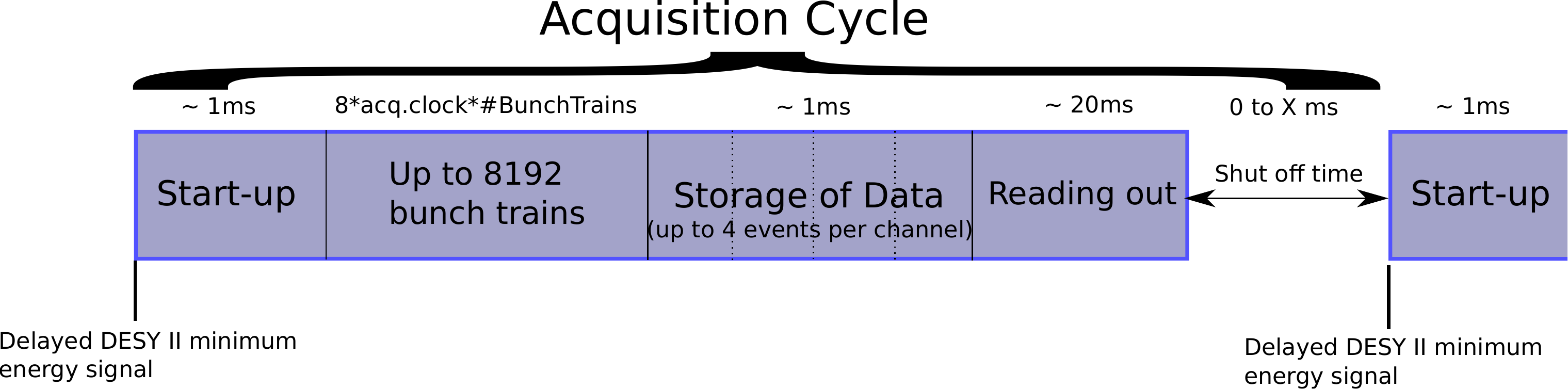}
				\caption{Working cycle of the KPiX readout.}
				\label{fig:acq.cycle}
			\end{center}
		\end{figure}

		The KPiX readout chip was designed by SLAC for the SiD ECAL detector and tracker system \cite{6551433}. The chip consists of a 1024 channel fully digital readout with a 13 bit ADC resolution and an internal $\SI{100}{MHz}$ clock. As the system was designed for an ILC environment, the chip performs power pulsing, i.e. the chip is only active for a period of time before turning off as seen in Figure \ref{fig:acq.cycle}. This cycle will in the following be referred to as the KPiX acquisition cycle (acq.cycle). The actual open period of the chip, the time where KPiX is capable of data taking, depends on the timing resolution set by the acquisition clock (acq. clock). The chip can store up to 8192 BunchTrains with the time of one BunchTrain, given by 
		\begin{equation} \label{eq:bunchclkcount}
			t_{\mathrm{BunchTrain}} = \mathrm{BunchClkCount}= 8 \cdot t_{\mathrm{acq.clock}}.
		\end{equation}		  
		  $t_{\mathrm{acq.clock}}$ can be chosen as multiples of $\SI{10}{ns}$ resulting from the $\SI{100}{MHz}$ clock. Measurement results shown here were done with a $t_{\mathrm{acq.clock}}$ of $\SI{320}{ns}$, resulting in an open period of
		\begin{equation} \label{eq:open_period}
			t_{\mathrm{open}} = 8\cdot t_{\mathrm{acq.clock}} \cdot\mathrm{BunchTrains} = 8 \cdot\SI{320}{ns} \cdot8192 = \SI{20.97}{ms}.
		\end{equation}
		In addition, to the open time of KPiX, after the data recording period the data is stored for about $\SI{1}{ms}$ before being read out by the Field Programmable Gate Array (FPGA) with the system being able to store up to 4 events per channel during one acquisition cycle as seen in Figure \ref{fig:acq.cycle}. Afterwards, all data is read out during a $\approx \SI{20}{ms}$ period after which KPiX turns off until either a certain amount of time passes or an external trigger signal is sent.
			
		KPiX can work in two operating modes, corresponding to two different triggering modes:
		\begin{enumerate}
			\item Self/internal triggering: If the charge generated within a channel rises above a user set threshold, the charge in the channel will be saved by KPiX.
			\item External triggering: An external signal is fed to KPiX which, at that moment, triggers every channel to store the recorded data it is.
		\end{enumerate}
 
\begin{figure} 
			\begin{center}
				\includegraphics[scale=0.37]{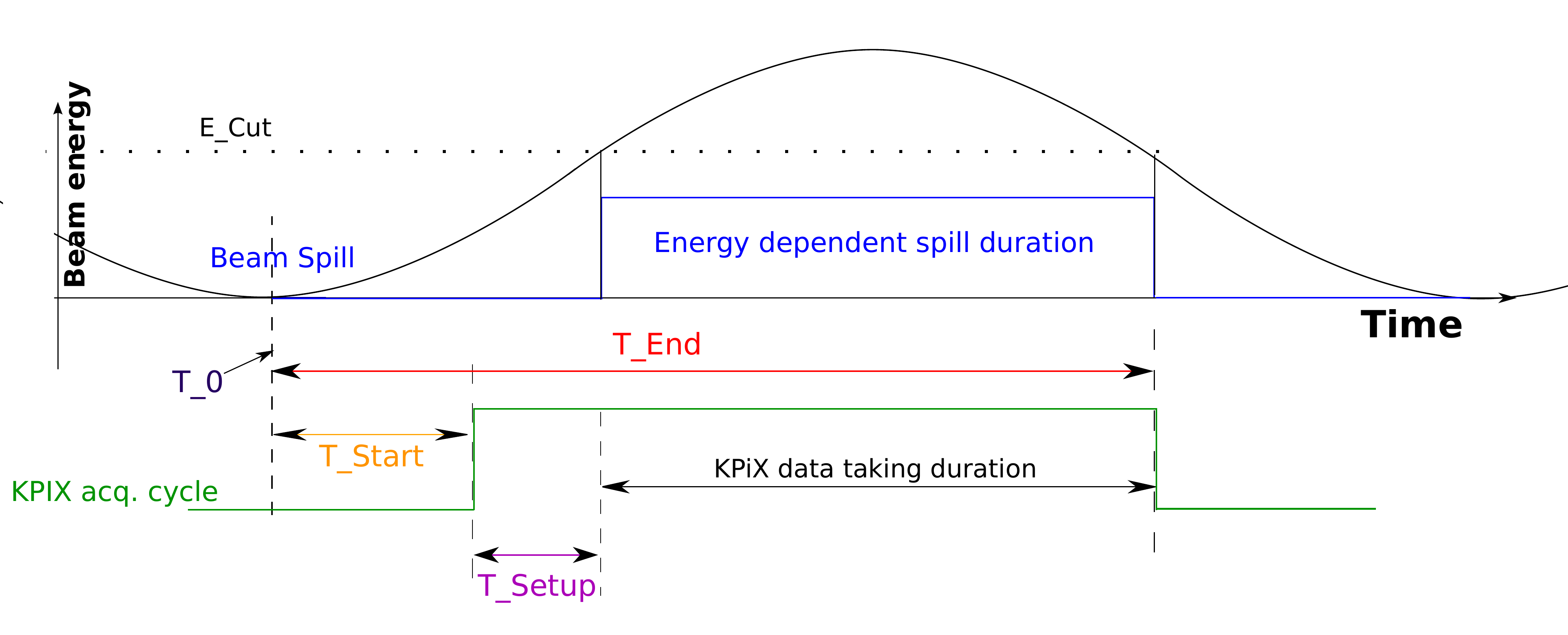}
				\caption{Sketch of the synchronization of KPiX with the DESY II cycle.}
				\label{fig:desy_synchronization}
			\end{center}
		\end{figure}

		As a result of the power pulsing, KPiX needs to be synchronized to the DESY II accelerator cycle to run with maximum efficiency.
		The DESY II cycle follows a sinusoidal energy curve that repeats every $\SI{80}{ms}$ with a new particle injection from the linac happening every $\SI{160}{ms}$. The spill of particles happens once the energy of the electrons within DESY II possess the chosen energy. 
		As a result, the beginning and duration of the particle spill depend on the chosen energy. Figure \ref{fig:desy_synchronization} shows a sketch of the method used to ensure that the system is only active when the energy dependent spill is happening. For this, the minimum energy signal \textcolor{purple}{$T_0$} is delayed by \textcolor{orange}{$\mathrm{T_{Start}}$} and used as a switch on signal for the system. After the setup period of \textcolor{violet}{$\mathrm{T_{Setup}} \approx \SI{1}{ms}$} KPiX is then active during the aforementioned open period, see equation: \eqref{eq:open_period}. The final system will in addition return a signal at \textcolor{red}{$\mathrm{T_{End}}$} to the Trigger Logic Unit (TLU) used for synchronization with the Device Under Test (DUT) once it has finished data taking. 
		
		\FloatBarrier
	\section{KPiX testing}
\begin{figure} 
			\begin{center}
				\includegraphics[width=0.9\textwidth]{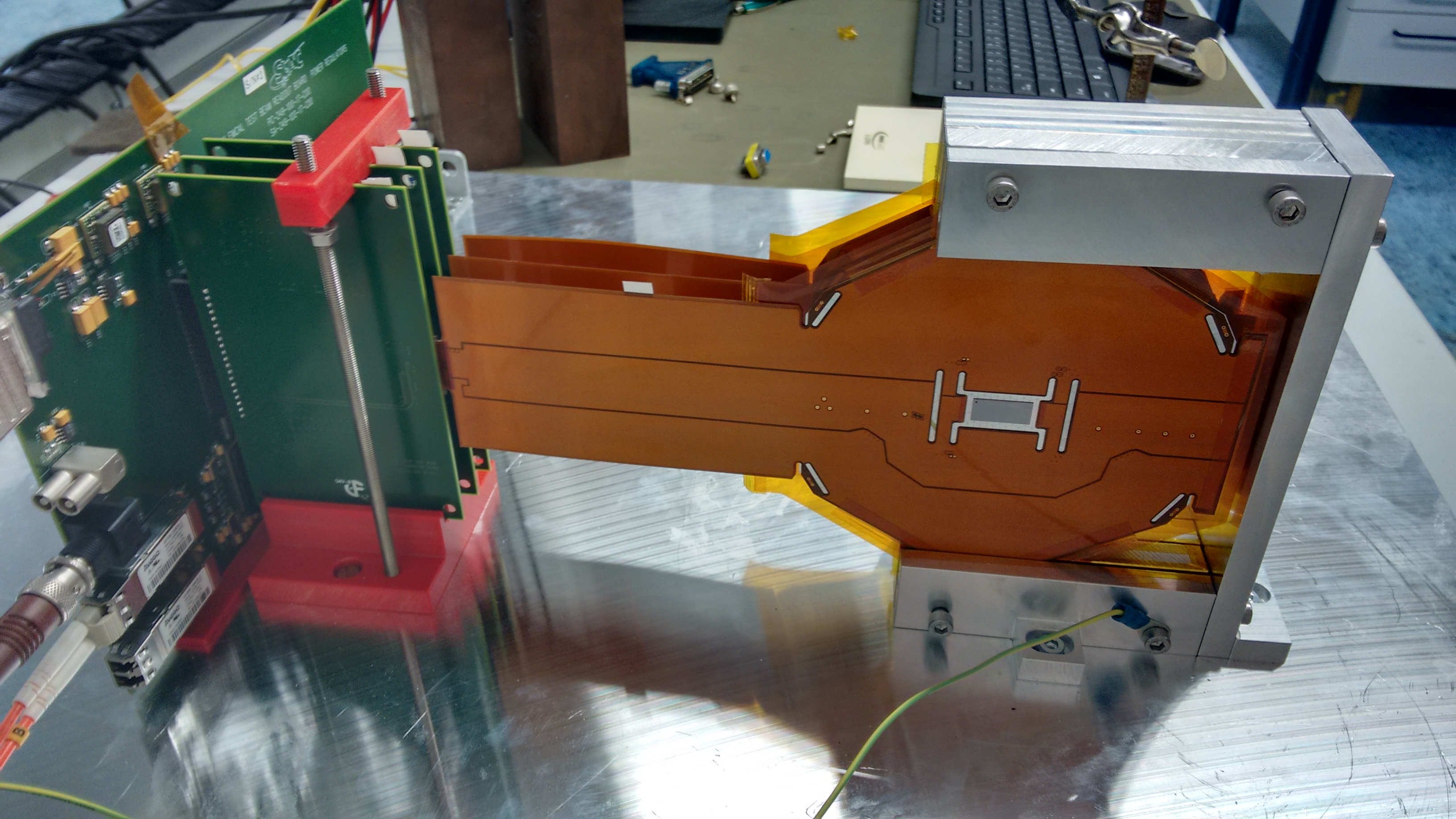}
				\caption{KPiX test setup consisting of a triple silicon ECAL stack, a readout board with a Xilinx Virtex-5 FPGA. Not shown here is the grounded aluminum dark box.}
				\label{fig:ecal_stack}
			\end{center}
		\end{figure}

		Before the final tracking sensors were available, SLAC kindly provided a test setup consisting of three silicon ECAL sensors that were set up at DESY. Each sensor consists of 1024 hexagonal pixels read out via a single bump bonded KPiX, see Figure \ref{fig:ecal_stack}. The setup was used to gain experience with the data readout and analysis.
		
\begin{figure} 
			\begin{center}
				\includegraphics[scale=0.3]{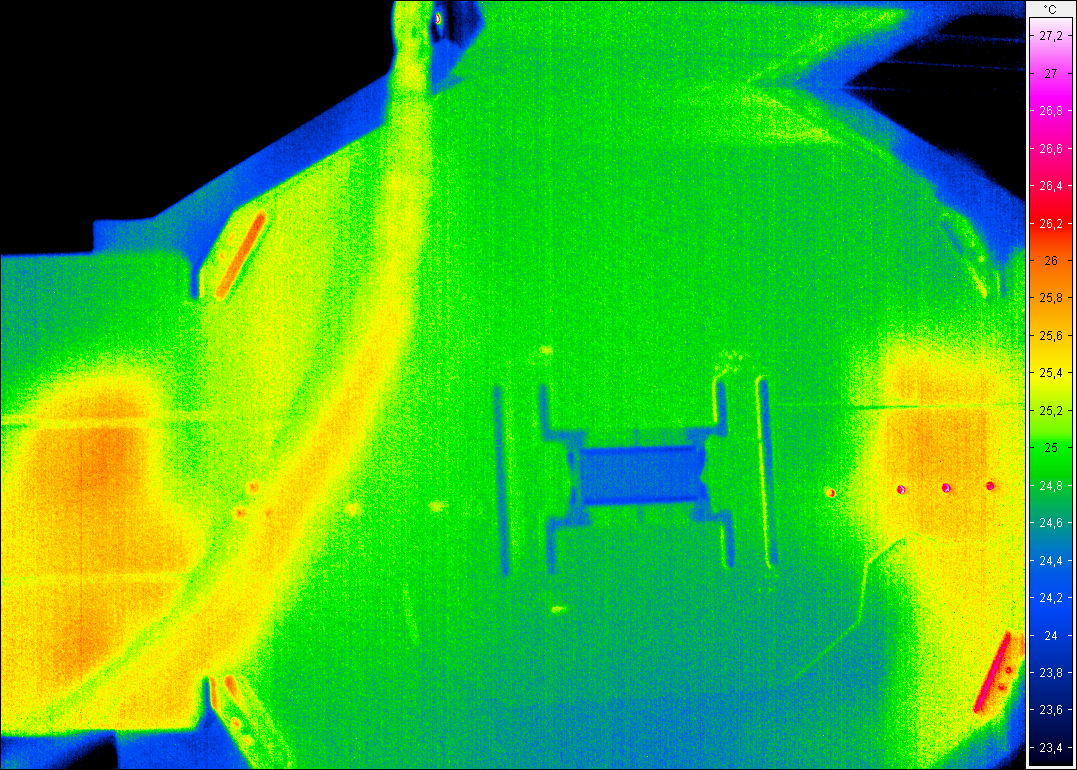}
				\caption{Infrared picture of the ECAL sensor during continuous running. The temperature scale ranges is from 23.4 to $\SI{27.2}{\celsius}$}
				\label{fig:ecal_infrared}
			\end{center}
		\end{figure}

		In addition to data readout and analysis, tests were conducted to measure the heat generated by the chip as it is important for the implementation of potential cooling systems which would, in consequence, increase the size of the support structure. A heat generation test was conducted by running KPiX in a mode as close to continuous running as possible to maximize the heat generation while using an infrared camera to observe the temperature of the sensor. 
		The measurements seen in Figure \ref{fig:ecal_infrared} show that the temperature of the sensor itself does not exceed $\approx \SI{25}{\celsius}$ anywhere except the small patches of high temperature that have been identified as reflections of external heat sources in the room. The chip and the surrounding area appear to be cooler than the rest of the sensor. This can be attributed to the higher emissivity of the silicon surface of the chip and the sensor compared to the Kapton layer that covers the rest of the sensor. Over the entire duration of the test no significant changes of the temperature were recorded.
		
		The data acquisition tests with KPiX were performed with the electron beam of the DESY II Test Beam Facility. The triple ECAL stack was positioned in the beam line with different configurations to test not only the synchronization with the DESY II energy cycle but also the triggering modes and the synchronization with external trigger signals. All measurements shown have been taken with an electron beam momentum of $p_{\mathrm{electron}} = \SI{4.0}{GeV}$.
		
\begin{figure} 
			\begin{center}
				\includegraphics[scale=0.6]{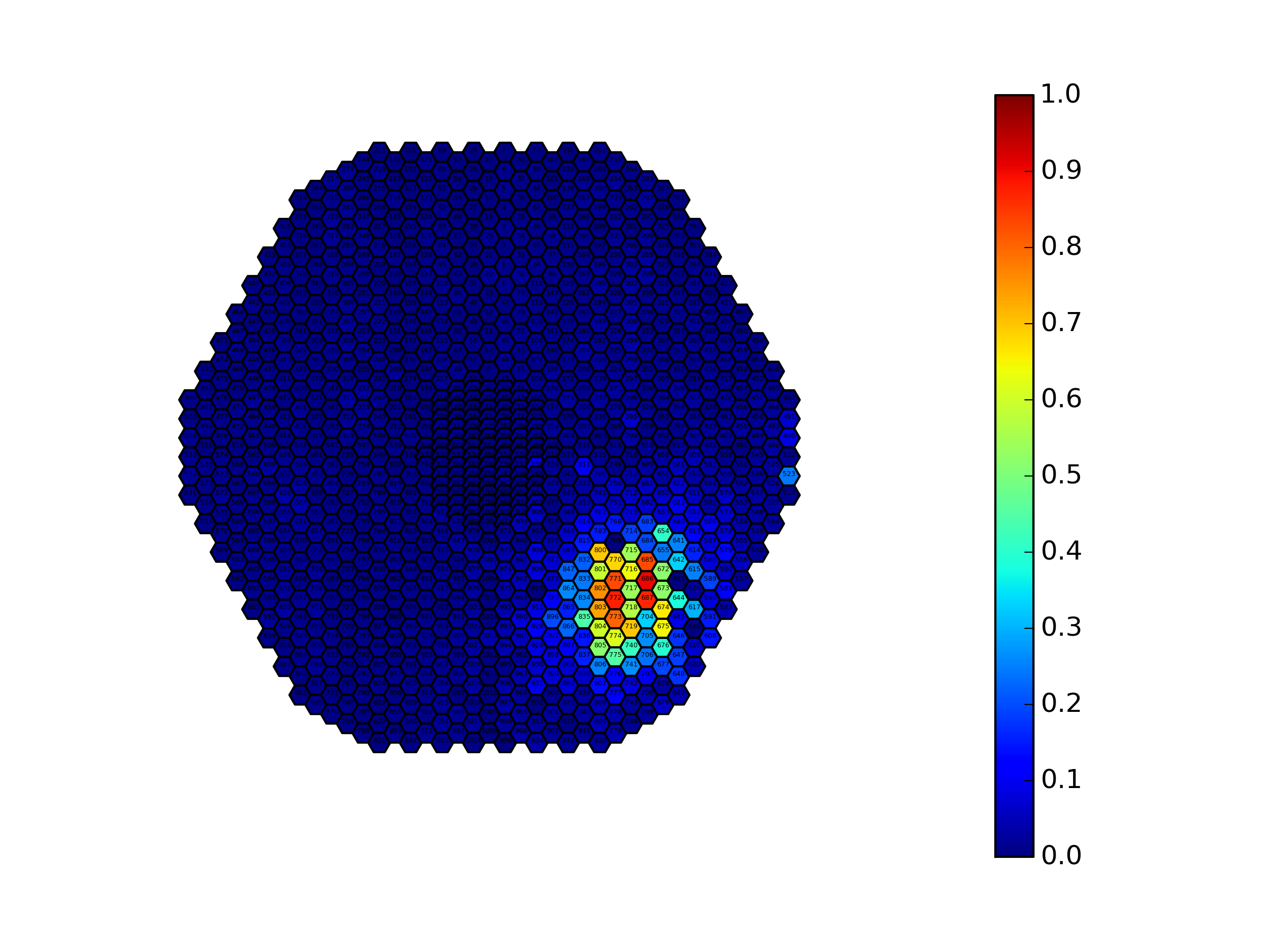}
				\caption{Number of entries normalized to the total number of acquisition cycles mapped onto the ECAl sensor.}
				\label{fig:ecal_mapped}
			\end{center}
		\end{figure}

		Figure \ref{fig:ecal_mapped} shows the amount of events registered mapped onto the ECAL sensor during a self-triggering run with 10000 acquisition cycles in total. One can see the beam spot in the bottom right corner of the map. The color code depicts the number of triggers normalized to the number of acquisition cycles.
		
\begin{figure} 
			\begin{center}
				\includegraphics[scale=0.7]{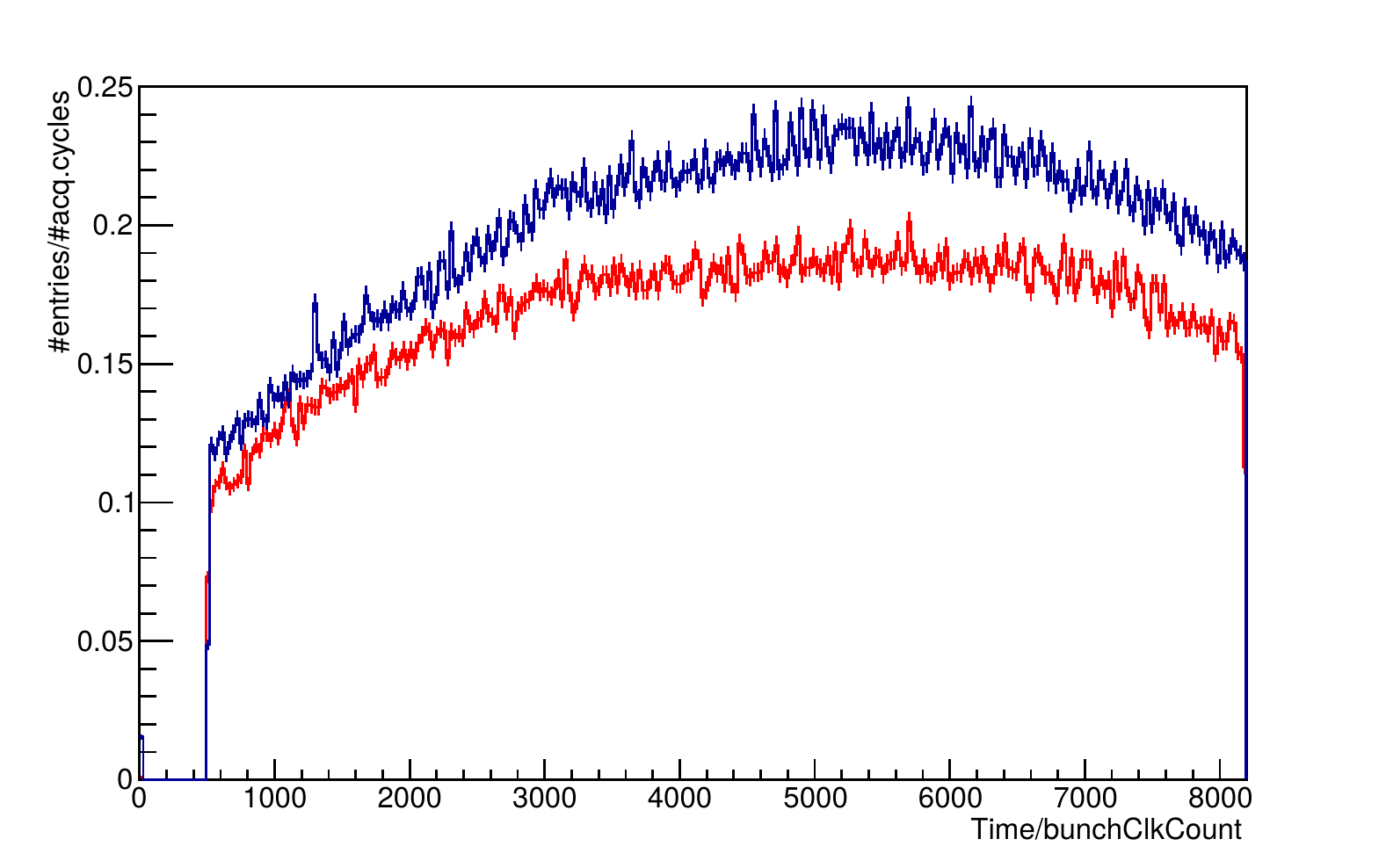}
				\caption{Time distribution of events sent by the external trigger scintillator in red with events registered by the silicon ECAL sensor in blue.}
				\label{fig:ext_int_time}
			\end{center}
		\end{figure}

\begin{figure} 
			\begin{center}
				\includegraphics[scale=0.7]{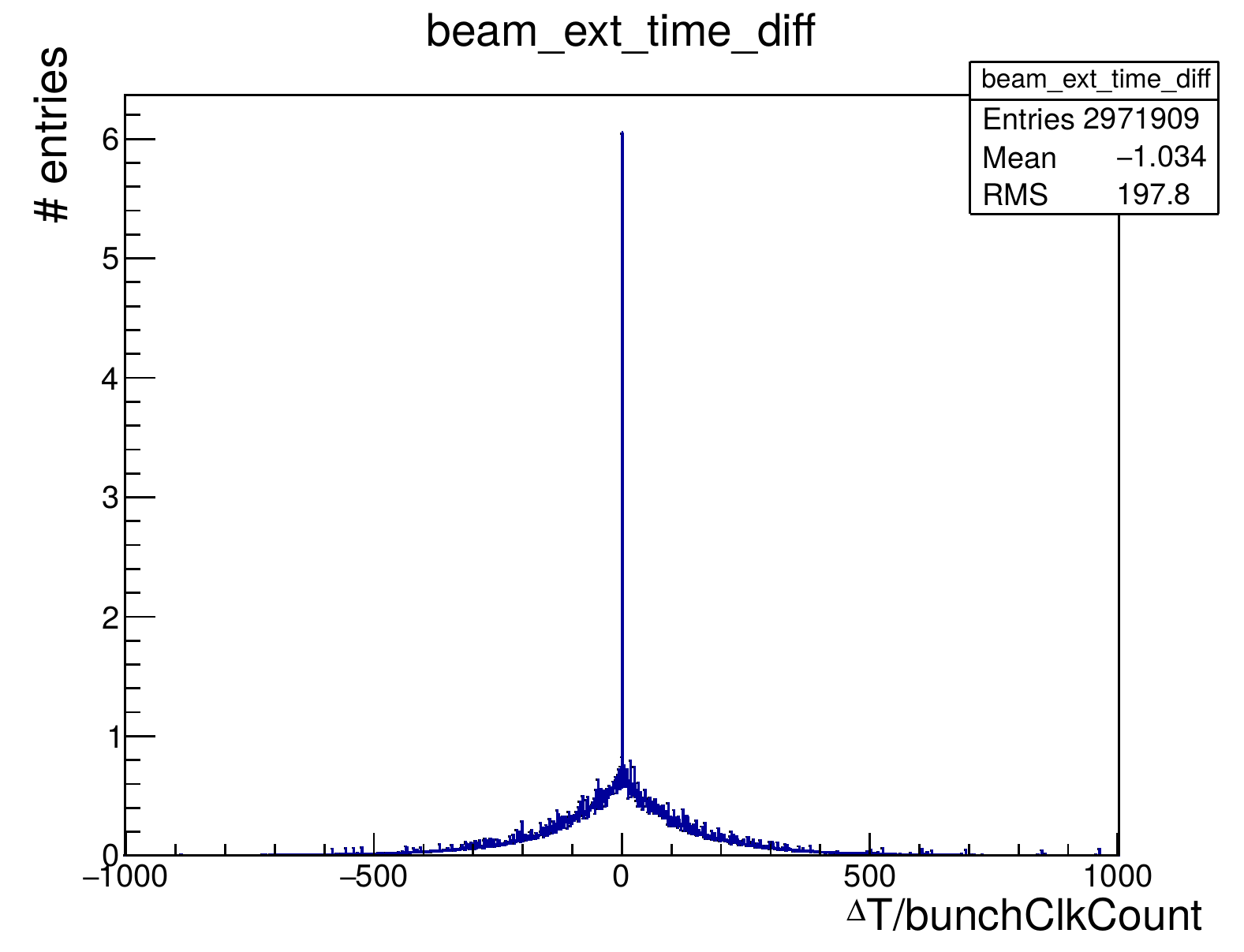}
				\caption{Time difference between internally triggered events and externally generated timestamps.}
				\label{fig:time_diff}
			\end{center}
		\end{figure}

		An external trigger signal generated by scintillator fingers located in the beam is sent to the KPiX DAQ board which is used to test external triggering and to serve as an external time measurement for noise reduction during self triggering. The results during the self triggering tests, summed up over 10000 acquisition cycles, are shown in Figure \ref{fig:ext_int_time}. The time information is given as multiples of the BunchClockCount$=\SI{2.56} {\micro s}$, see equation: \eqref{eq:bunchclkcount}. 
		The increase and subsequent decrease in the number of registered events over the open time of KPiX corresponds to the number of particles with a momentum higher than $E_{\mathrm{electron}} = \SI{4.0}{GeV}$. 
		 The empty bins from 0 to 500 were manually suppressed, as one of the chips showed a high noise sensitivity shortly after startup.
		The external time stamps are recorded as a list of time stamps per acquisition cycle in the data file.  In order to match the external time stamps to triggered channels, every single channel is compared to the list of external time stamps and the difference $\Delta T = t_{\mathrm{internal}} - t_{\mathrm{external}}$ is calculated looking for the minimum $\Delta T$. The minimum is then filled into the histogram in Figure \ref{fig:time_diff}. The large peak in the Figure is at 2 BunchClkCounts, equating to approximately $\SI{5.12}{\micro s}$, showing there is a general time offset between the external and internal time stamps. A cut of $0 \leq \Delta T \leq 3$ in units of the BunchClkCount has been applied around this peak to reduce the events triggered by noise in the system.
		
		\FloatBarrier
	\section{The Final System}
		As a general infrastructure upgrade for the DESY II Test Beam Facility, the system needs to be robust and easy to operate by external user groups. The system should in addition be flexible to be used outside of the PCMAG in a similar manner to the DATURA/DURANTA telescopes. Therefore, a modular system is produced that allows the sensors to be easily uninstalled from the PCMAG while ensuring the necessary protection of the sensors. 
		
\begin{figure} 
			\begin{center}
				\includegraphics[width=0.7\textwidth]{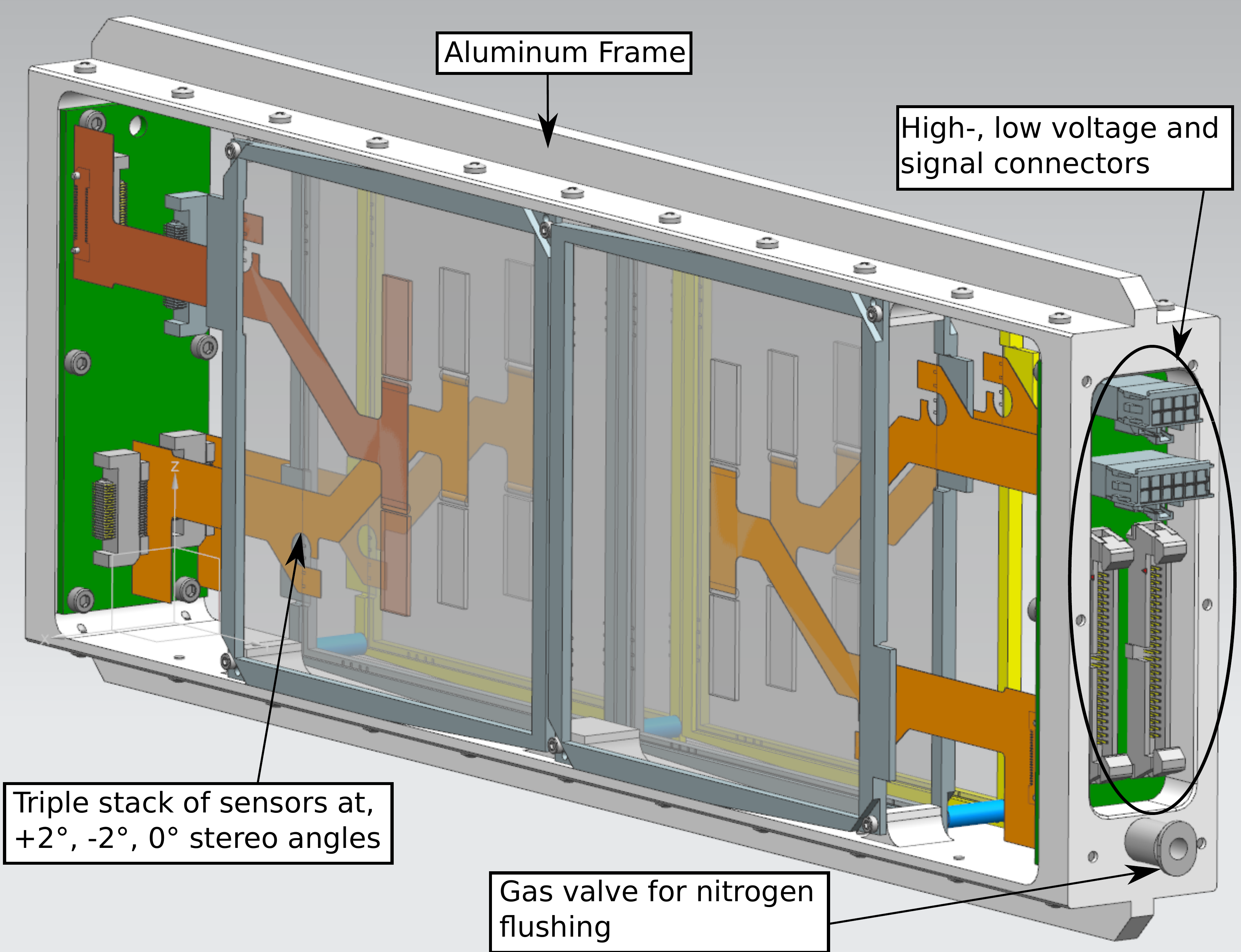}
				\caption{CAD drawing of the telescope cassette, of which one will be placed on each side of the DUT.}
				\label{fig:CAD_cassette}
			\end{center}
		\end{figure}

\begin{figure} 
			\begin{center}
				\includegraphics[width=0.7\textwidth]{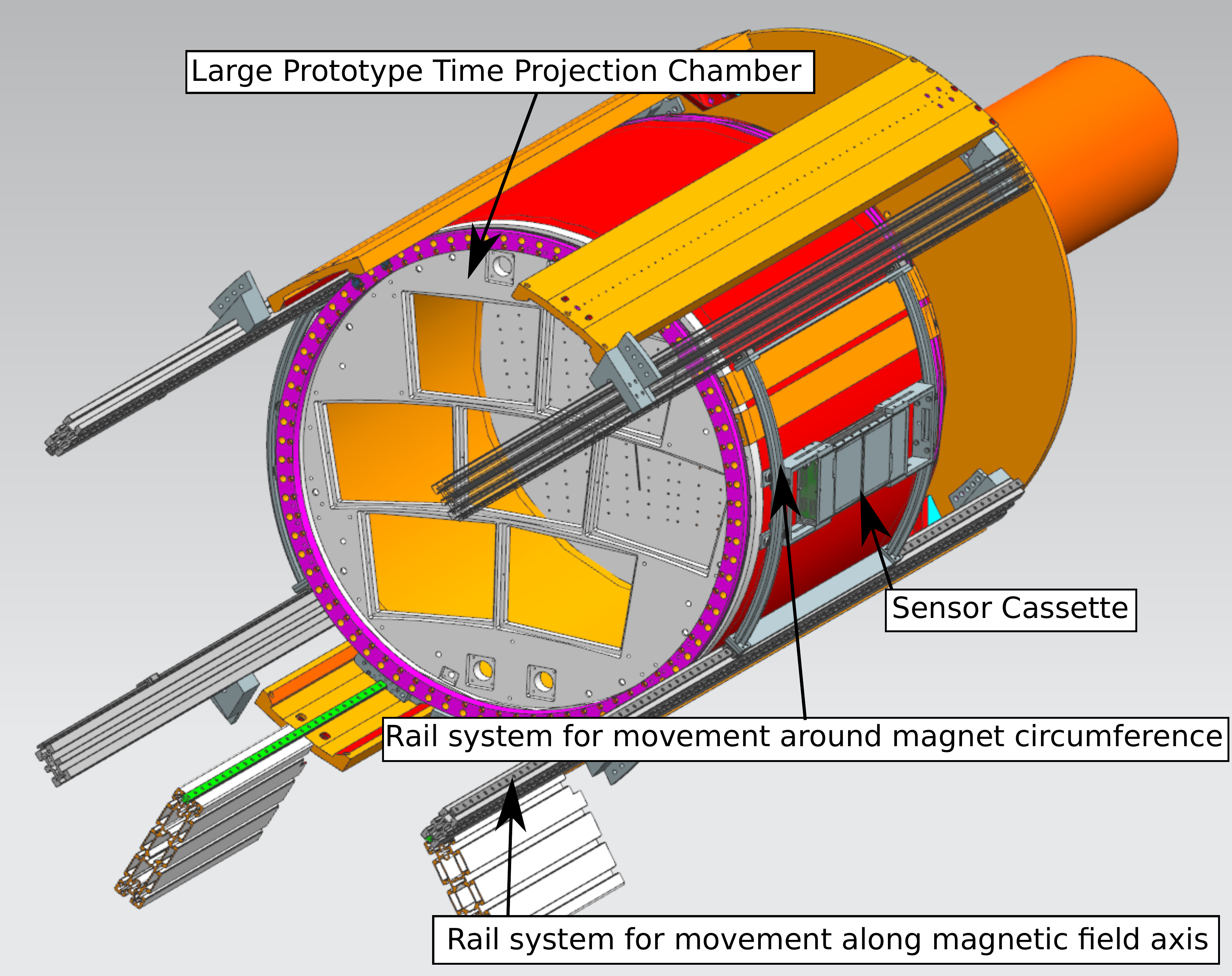}
				\caption{CAD drawing of the silicon telescope within the PCMAG together with the Large Prototype TPC. The cassette in the picture is shown in Figure \ref{fig:CAD_cassette}.}
				\label{fig:CAD_magnet_structure}
			\end{center}
		\end{figure}

		The final system will consist of separate substructures, i.e. two cassettes of which each consists of, see Figure \ref{fig:CAD_cassette}: 
		
		\begin{itemize}
			\item An aluminum frame with carbon fiber windows on both sides to provide mechanical stability, protection of the sensors, light tightness and electrical shielding.
			\item Electronic boards inside the cassette, equipped with noise filters and linear regulators, for electronic routing, low voltage supply for the KPiX and electronic connection to the outside.
			\item Two side by side stacks of 3 strip sensors with a distance of $\SI{15}{mm}$ between each sensor, $+\SI{2}{\degree}$, $-\SI{2}{\degree}$ and $\SI{0}{\degree}$ stereo angles to avoid hodoscope ambiguity and provide large coverage with the needed resolution along the axis of the magnetic field.
			\item A gas valve used for nitrogen flushing to keep the humidity inside the cassette at a minimum.
		\end{itemize}
		As well as a mounting system within the PCMAG, see Figure \ref{fig:CAD_magnet_structure}:
		\begin{itemize}
			\item Two rails on each side of the inside of the PCMAG to support the cassettes and allow for movement in the z direction along the magnetic field axis.
			\item A small cart mounted on the two rails on one side holds the cassette and allows for movement of the cassette along the circumference of the PCMAG.
		\end{itemize}
		It is planned to implement a position measurement system to precisely determine and record the position of the cassette in the PCMAG.
		\FloatBarrier
	\section{Next steps}
		The next step after the bump bonding at Fraunhofer IZM will be to repeat the test of the electrical properties of the sensors. Following this, the Kapton flex cable will be glued onto the silicon sensor using Araldite 2011 glue. This cable provides, the power supply of the chip, the bias voltage of the sensor and the signal readout. The electronic cassette boards and the DAQ board are designed by SLAC. The boards are expected to be finished in early 2018. The cassette is produced by DESY and will be completed in the middle of February, the rail system production will follow in March.
		
		After the full assembly of the sensors in the cassette, the tracker system will be tested at the DESY II Test Beam Facility. In addition, the DATURA/DURANTA telescopes within the PCMAG will be used for the commissioning of the telescope and for the final alignment of the telescope system.
		In addition, there is work ongoing on integration of the system into the existing EUDAQ framework to facilitate the use as well as to provide a common data taking and analysis framework similar to the DATURA/DURANTA telescope.
		The deliverable for this AIDA2020 project is in April 2017 after which the system will be available to external users.
		\FloatBarrier
	\section{Acknowledgments}
		This project has received funding from the European Union’s Horizon 2020 Research and Innovation programme under Grant Agreement no. 654168.
		
		Special thanks go to M. Breidenbach, D. Freytag and B.A.Reese from the SLAC Particle Physics and Astrophysics group for their support with the KPiX readout and ECAL test setup. 
		
		In addition, thanks go to D. Cussans and P. Baesso from the Bristol University Particle Physics group for their help concerning the data synchronization of KPiX with the DESY II Test Beam Facility.
	\bibliographystyle{ieeetr}
	\bibliography{/home/kraemeru/Documents/Presentations/LCWS2017/proceeding}

\begin{thebibliography}{10}

\bibitem{testbeamwebsite}
``{DESY II} {Test} {Beam} {Facility} website.''
  \url{http://particle-physics.desy.de/test_beams_at_desy/beam_generation/index_ger.html}.
\newblock Accessed: 2018-01-18.

\bibitem{pcmag:magnet}
A.~Yamamoto, K.~Anraku, R.~Golden, T.~Haga, Y.~{Higashi}, M.~Imori, S.~Inaba,
  B.~Kimbell, N.~{Kimura}, and Y.~Makida, ``{Balloon-borne experiment with a
  superconducting solenoidal magnet spectrometer},'' {\em {Advances in Space
  Research}}, vol.~14, p.~2, feb 1994.

\bibitem{pcmag:fieldmeas}
J.~Alozy, F.~Bergsma, F.~Formenti, {\em et~al.}, ``{First Version of the PCMAG
  Field Map},'' {\em {Eudet-Memo}}, vol.~{2007-51}, 2007.

\bibitem{pcmag:fieldana}
C.~Grefe, ``{Magnetic Field Map for a Large TPC Prototype},'' Master's thesis,
  {Universit\"at Hamburg}, {June} 2008.
\newblock {DESY-THESIS-2008-052}.

\bibitem{EudetTel2016}
H.~Jansen, S.~Spannagel, J.~Behr, {\em et~al.}, ``Performance of the
  {E}{U}{D}{E}{T}-type beam telescopes,'' {\em EPJ Techniques and
  Instrumentation}, 2016.

\bibitem{EUDETMemo2009}
D.~Cussans, ``Description of the {J}{R}{A}1 {Trigger} {Logic} {Unit}
  ({T}{L}{U}), v0.2c, {Executive} {Summary},'' tech. rep., Bristol, 2009.

\bibitem{tpc2004}
S.~Klein, ``The time projection chamber turns 25,'' {\em CERN COURIER}, 2004.

\bibitem{Mueller:2016exq}
F.~J. M{\"u}ller, {\em {Development of a Triple GEM Readout Module for a Time
  Projection Chamber \& Measurement Accuracies of Hadronic Higgs Branching
  Fractions in $\nu\nu$H at a 350 GeV ILC}}.
\newblock PhD thesis, Hamburg U., Hamburg, 2016.

\bibitem{ILD2013}
H.~Abramowicz {\em et~al.}, ``{The {International} {Linear} {Collider}
  {Technical} {Design} {Report} - {Volume} 4: {Detectors}},'' 2013.

\bibitem{ILC2013}
T.~Behnke, J.~E. Brau, B.~Foster, J.~Fuster, M.~Harrison, J.~M. Paterson,
  M.~Peskin, M.~Stanitzki, N.~Walker, and H.~Yamamoto, ``{The {International}
  {Linear} {Collider} {Technical} {Design} {Report} - {Volume} 1: {Executive}
  {Summary}},'' 2013.

\bibitem{VERBITSKAYA200347}
E.~Verbitskaya {\em et~al.}, ``The effect of charge collection recovery in
  silicon p–n junction detectors irradiated by different particles,'' {\em
  Nuclear Instruments and Methods in Physics Research Section A: Accelerators,
  Spectrometers, Detectors and Associated Equipment}, vol.~514, no.~1, pp.~47
  -- 61, 2003.
\newblock Proceedings of the 4th International Conference on Radiation Effects
  on Semiconductor Materials, Detectors and Devices.

\bibitem{6551433}
J.~Brau, M.~Breidenbach, A.~Dragone, G.~Fields, R.~Frey, D.~Freytag,
  M.~Freytag, C.~Gallagher, G.~Haller, R.~Herbst, B.~Holbrook, R.~Lander,
  A.~Moskaleva, C.~Neher, T.~Nelson, S.~Schier, B.~Schumm, D.~Strom,
  M.~Tripathi, and M.~Woods, ``K{P}i{X} - {A} 1,024 channel readout
  {A}{S}{I}{C} for the {I}{L}{C},'' in {\em 2012 IEEE Nuclear Science Symposium
  and Medical Imaging Conference Record (NSS/MIC)}, pp.~1857--1860, Oct 2012.

\end{thebibliography}
\end{document}